\documentclass[aps,prl,twocolumn,superscriptaddress,showpacs]{revtex4}
\usepackage{graphicx}

\begin{document}


\title{Limits on the spin-dependent WIMP-nucleon cross-sections\\
from the first science run of the ZEPLIN-III experiment}

\author{V.~N.~Lebedenko\footnote{Deceased}}
\affiliation{Blackett Laboratory, Imperial College London, UK}
\author{H.~M.~Ara\'ujo\footnote{Corresponding author; Email: h.araujo@imperial.ac.uk}}
\affiliation{Blackett Laboratory, Imperial College London, UK}
\affiliation{Particle Physics Department, Rutherford Appleton Laboratory, Chilton, UK}
\author{E.~J.~Barnes}
\affiliation {School of Physics and Astronomy, University of Edinburgh, UK}
\author{A. Bewick}
\affiliation{Blackett Laboratory, Imperial College London, UK}
\author{R.~Cashmore}
\affiliation{Brasenose College, University of Oxford, UK}
\author{V.~Chepel}
\affiliation{LIP--Coimbra \& Department of Physics of the University of Coimbra, Portugal}
\author{D.~Davidge}
\affiliation{Blackett Laboratory, Imperial College London, UK}
\author{J.~Dawson}
\affiliation{Blackett Laboratory, Imperial College London, UK}
\author{T.~Durkin}
\affiliation{Particle Physics Department, Rutherford Appleton Laboratory, Chilton, UK}
\author{B.~Edwards}
\affiliation{Blackett Laboratory, Imperial College London, UK}
\affiliation{Particle Physics Department, Rutherford Appleton Laboratory, Chilton, UK}
\author{C.~Ghag}
\affiliation{School of Physics and Astronomy, University of Edinburgh, UK}
\author{V.~Graffagnino}
\affiliation{Particle Physics Department, Rutherford Appleton Laboratory, Chilton, UK}
\author{M.~Horn}
\affiliation{Blackett Laboratory, Imperial College London, UK}
\author{A.~S.~Howard}
\affiliation{Blackett Laboratory, Imperial College London, UK}
\author{A.~J.~Hughes}
\affiliation{Particle Physics Department, Rutherford Appleton Laboratory, Chilton, UK}
\author{W.~G.~Jones}
\affiliation{Blackett Laboratory, Imperial College London, UK}
\author{M.~Joshi}
\affiliation{Blackett Laboratory, Imperial College London, UK}
\author{G.~E.~Kalmus}
\affiliation{Particle Physics Department, Rutherford Appleton Laboratory, Chilton, UK}
\author{A.~G.~Kovalenko}
\affiliation{Institute for Theoretical and Experimental Physics, Moscow, Russia}
\author{A.~Lindote}
\affiliation{LIP--Coimbra \& Department of Physics of the University of Coimbra, Portugal}
\author{I.~Liubarsky}
\affiliation{Blackett Laboratory, Imperial College London, UK}
\author{M.~I.~Lopes}
\affiliation{LIP--Coimbra \& Department of Physics of the University of Coimbra, Portugal}
\author{R.~L\"{u}scher}
\affiliation{Particle Physics Department, Rutherford Appleton Laboratory, Chilton, UK}
\author{K.~Lyons}
\affiliation{Blackett Laboratory, Imperial College London, UK}
\author{P.~Majewski}
\affiliation{Particle Physics Department, Rutherford Appleton Laboratory, Chilton, UK}
\author{A.~StJ.~Murphy}
\affiliation {School of Physics and Astronomy, University of Edinburgh, UK}
\author{F.~Neves}
\affiliation{LIP--Coimbra \& Department of Physics of the University of Coimbra, Portugal}
\author{J.~Pinto da Cunha}
\affiliation{LIP--Coimbra \& Department of Physics of the University of Coimbra, Portugal}
\author{R.~Preece}
\affiliation{Particle Physics Department, Rutherford Appleton Laboratory, Chilton, UK}
\author{J.~J.~Quenby}
\affiliation{Blackett Laboratory, Imperial College London, UK}
\author{P.~R.~Scovell}
\affiliation {School of Physics and Astronomy, University of Edinburgh, UK}
\author{C.~Silva}
\affiliation{LIP--Coimbra \& Department of Physics of the University of Coimbra, Portugal}
\author{V.~N.~Solovov}
\affiliation{LIP--Coimbra \& Department of Physics of the University of Coimbra, Portugal}
\author{N.~J.~T.~Smith}
\affiliation{Particle Physics Department, Rutherford Appleton Laboratory, Chilton, UK}
\author{P.~F.~Smith}
\affiliation{Particle Physics Department, Rutherford Appleton Laboratory, Chilton, UK}
\author{V.~N.~Stekhanov}
\affiliation{Institute for Theoretical and Experimental Physics, Moscow, Russia}
\author{T.~J.~Sumner}
\affiliation{Blackett Laboratory, Imperial College London, UK}
\author{C.~Thorne}
\affiliation{Blackett Laboratory, Imperial College London, UK}
\author{R.~J.~Walker}
\affiliation{Blackett Laboratory, Imperial College London, UK}

\collaboration{ZEPLIN-III}

\date{\today}

\begin{abstract}

We present new experimental constraints on the WIMP-nucleon
spin-dependent elastic cross-sections using data from the first
science run of ZEPLIN-III, a two-phase xenon experiment searching for
galactic dark matter WIMPs based at the Boulby mine. Analysis of
$\sim$450~kg$\cdot$days fiducial exposure revealed a most likely
signal of zero events, leading to a 90\%-confidence upper limit on the
pure WIMP-neutron cross-section of $\sigma_n\!=\!1.8\!\times\!
10^{-2}$~pb at 55~GeV/$c^2$ WIMP mass. Recent calculations of the
nuclear spin structure based on the Bonn~CD nucleon-nucleon potential
were used for the odd-neutron isotopes $^{129}$Xe and
$^{131}$Xe. These indicate that the sensitivity of xenon targets to
the spin-dependent WIMP-proton interaction is much lower than implied
by previous calculations, whereas the WIMP-neutron sensitivity is
impaired only by a factor of $\sim$2.

\end{abstract}


\maketitle

ZEPLIN-III has recently completed its first run at the Boulby
Underground Laboratory (UK) searching for weakly interacting massive
particles (WIMPs), which have been proposed to explain the
non-baryonic cold dark matter in the Universe. In several
supersymmetry (SUSY) extensions to the Standard Model of particle
physics, the lightest supersymmetric particle (LSP) is stable and its
relic abundance could account for the matter content observed today
\cite{jungmann96}. In most SUSY scenarios this particle is the
neutralino, $\tilde{\chi}$, which is a linear combination of the SUSY
partners of the electroweak gauge bosons (gauginos) and Higgs bosons
(higgsinos). Neutralinos are expected to scatter elastically from
ordinary matter, producing low-energy nuclear recoils, predominantly
through scalar (or spin-independent, SI) interactions. This is
especially so for heavier elements ($A\!\agt\!30$, with $A$ the number
of nucleons), for which the scattering involves the entire nucleus
rather than individual nucleons. This coherence enhancement of the
scalar term in the interaction cross-section is proportional to
$A^2$. Our SI result excludes a WIMP-nucleon cross-section above
7.7$\times$10$^{-8}$~pb for 55~GeV/$c^2$ mass with 90\%
confidence~\cite{lebedenko08a}.

Nonetheless, axial-vector (or spin-dependent, SD) interactions could
dominate in some SUSY scenarios in regions where the SI term is
suppressed~\cite{bednyakov94,bednyakov04}. In addition, interpretation
of the DAMA annual modulation \cite{barnabei00,barnabei08} in terms of
nuclear recoils caused by a dominant SD interaction \cite{savage04}
has remained viable for WIMP masses below $\sim$15 GeV until
recently. It is therefore important to continue pursuing the search in
the SD channel. Xenon targets have good sensitivity to the
WIMP-neutron interaction since approximately half of natural abundance
consists of the odd-neutron isotopes $^{129}$Xe and $^{131}$Xe (the
proton interaction is highly suppressed for these isotopes, and
neither is likely in the even-even nuclei). 

It is also possible that a SD inelastic interaction with these nuclei
could be significant for relatively heavy WIMPs
\cite{belli96,toivanen08}. Deexcitation of low-lying nuclear states in
$^{129}$Xe and $^{131}$Xe (emitting 39.6~keV and 80.2~keV
$\gamma$-rays, respectively) provides a detection mechanism with low
effective energy threshold, since the $\gamma$-rays can be efficiently
detected. Although neutrons also scatter inelastically producing the
same signatures, differences in spatial distribution between signal
and background might be exploited in very large detectors -- assuming
that the dominant $\gamma$-ray backgrounds could be suitably
mitigated. Most direct search experiments are still focused on
detecting nuclear recoils down to $\sim$keV energies; this is so in
the case described in this Letter.


WIMP search experiments propose to measure the total WIMP-nucleus
elastic cross-section, $\sigma_A$. Assuming dominance of either the SI
or the SD term, one can write:
\begin{equation}
  \sigma_A = 4 G_F^2 \mu_A^2 C_A \, ,
  \label{sigma_A}
\end{equation}
where $G_F$ is the Fermi weak-coupling constant, $\mu_A$ is the
reduced mass of the WIMP-nucleus system, and $C_A$ is an enhancement
factor which depends on the type of interaction and, possibly, the
WIMP composition. For the SI case, $C_A\!\propto\!A^2$ (i.e. $C_A$ is
model-independent for Majorana WIMPs). For a dominant SD case, $C_A$
involves the total nuclear spin $J$ instead:
\begin{equation}
  C_A = \frac{8}{\pi} 
  \left(  a_p\langle S_p \rangle + a_n\langle S_n \rangle \right)^2
  \frac{J+1}{J} \, ,
  \label{C_A}
\end{equation}
where $\langle S_{p,n} \rangle$ are the expectation values for the
proton or neutron spins averaged over the nucleus and $a_{p,n}$ are
the effective WIMP-nucleon coupling constants. The latter depend on
the WIMP particle content (e.g.~more higgsino- or gaugino-like) as
well as the quark spin distributions inside the nucleons.

Two spin distribution calculations were adopted, both based on the
nuclear shell model but using different nucleon-nucleon ($nn$)
potentials (see, e.g., reviews \cite{hjort95,coraggio09}). Their
ability to reproduce experimental measurements of the magnetic moment
is the standard benchmark, as this observable is reasonably similar to
the WIMP-nucleus scattering matrix element. For comparison with other
Xe experiments we use spin structures with the Bonn~A potential for
both isotopes \cite{ressell97}. Agreement of the magnetic moment (using
effective $g$-factors) is quite reasonable: within 19\% and 8\% for
$^{129}$Xe and $^{131}$Xe, respectively. In addition, new calculations
based on the Bonn-CD G-matrix have become available which improve on
these figures significantly (to 1\% and 2\%) for both isotopes
\cite{toivanen08}. The spin expectation values $\langle S_{p,n}
\rangle$ are given in Table~I for both cases.

In order to compare (\ref{sigma_A}) with an experimental result a
nuclear form factor, $F^2(q)$, must be assumed at momentum transfer
$q\!>\!0$ to account for finite nuclear size. Although cross-sections
are conventionally given in the limit of zero momentum transfer (the
`standard' cross-section), $F^2(q)$ must be factored into
(\ref{sigma_A}) to describe the interaction with a particular
nucleus. The definition of $C_A$ in (\ref{C_A}) justifies the use of a
normalised form factor $F^2(q)$=$S(q)/S(0)$, where $S(q)$ folds the
spin structure functions $S_{jj\prime}(q)$ with an arbitrary
neutralino composition in the isospin convention. The related
functions $F_{\rho\rho\prime}(u)$ derived with Bonn
CD~\cite{toivanen08} were converted into the same $S_{jj\prime}(q)$
normalisation used with Bonn~A~\cite{ressell97} by employing Eq.~(18)
in Ref.~\cite{holmlund04} and then parametrised with a 6$^{th}$-order
polynomial for low $q$. As noted in Ref.~\cite{toivanen08}, the new
calculation gives smaller spin structures than those found for simpler
models: by a factor of $\sim$2 for Xe$^{129}$ and $\sim$4 for
Xe$^{131}$ at low $q$. More significantly, the spin factors are also
smaller, in particular $\langle S_{p}\rangle$.

\begin{table}
\caption{Xe isotope parameters: nuclear spin $J$, isotopic fraction
(nat. abundance), effective exposure and spin factors $\langle S_{n,p}
\rangle$ with Bonn~A~\cite{ressell97} and Bonn~CD~\cite{toivanen08}
$nn$ potentials.}
\begin{ruledtabular}
\begin{tabular}{lcccccccc}
&&&&&\multicolumn{2}{c}{Bonn A}&\multicolumn{2}{c}{Bonn CD}\\
& $J$ & \% & (n.a.) & kg$\cdot$days & $\langle S_n \rangle$ & $\langle S_p \rangle$ & $\langle S_n \rangle$ & $\langle S_p \rangle$\\
\hline
$^{129}$Xe & 1/2 & 29.5 & (26.4) & 37.7 & 0.359 & 0.028 & 0.273 & -0.0019 \\
$^{131}$Xe & 3/2 & 23.7 & (21.2) & 33.7 & -0.227 & -0.009 & -0.125 & -0.00069 
\end{tabular}
\end{ruledtabular}
\end{table}

Calculating the cross-section per nucleon allows comparison of
different target materials and with theoretical model predictions,
which are usually computed for free protons and neutrons. This
conversion is not straightforward for the SD case since the
cross-section has contributions from both proton and neutron terms, as
indicated explicitly in (\ref{C_A}). In addition, $F^2(q)$ is
similarly contaminated by both couplings. A simple approach exists
that allows a straightforward calculation in a model-independent way
by assuming that WIMPs couple predominantly to protons {\em or}
neutrons \cite{tovey00}. By setting, in turn, $a_n$=0 and $a_p$=0, no
assumption is required as to the neutralino composition, either
explicitly in the standard cross-section or in the form factor. In
this instance one may write, in the limit of zero momentum transfer:
\begin{equation}
  \frac{\sigma_{p,n}}{\sigma_A} = \frac{3}{4}\frac{\mu_{p,n}^2}{\mu_A^2}
  \frac{1}{\langle S_{p,n}\rangle^2}\frac{J}{J+1} \, ,
  \label{sigma_n}
\end{equation}
where $\mu_{p,n}$ is the WIMP-nucleon reduced mass. Once $\sigma_A$ is
obtained from experimental limits on the allowed nuclear recoil
spectrum, (\ref{sigma_n}) is used to calculate the corresponding SD
cross-section for each isotope, $\sigma_{p,n}^{129}$ and
$\sigma_{p,n}^{131}$. These are combined to obtain the total
cross-sections: $(\sigma_{p,n})^{-1} = (\sigma_{p,n}^{129})^{-1} +
(\sigma_{p,n}^{131})^{-1}$.


ZEPLIN-III is a two-phase xenon time projection chamber operating
1100~m underground at the Boulby mine. It is able to discriminate
between nuclear recoil signals and the more prevalent electron recoil
background by measuring both the scintillation light (S1) and the
ionisation charge (S2) produced by interactions in its 12~kg liquid
xenon target. 3D reconstruction of the interaction vertex, a strong
electric field in the liquid (3.9~kV/cm) and a geometry which avoids
surfaces near a central fiducial region allow powerful discrimination
down to low energies. Data analysis procedures are detailed with our
SI result~\cite{lebedenko08a} and briefly summarised here. The
instrument is further described in Refs.~\cite{araujo06,akimov07}.

A WIMP acceptance region was defined in the range [10.7,30.2]~keV
recoil energy and [$\mu$--$2\sigma$,$\mu$] in the $\log_{10}$(S2/S1)
discrimination parameter, where $\mu$ and $\sigma$ describe the means
and standard deviations of log-normal distributions fitted to the
nuclear recoil population produced by elastic neutron scattering over
that energy range. An effective exposure of 128.7~kg$\cdot$days was
accumulated during the 83-day science run between 27$^{\rm th}$~Feb
and 20$^{\rm th}$ May 2008. This exposure is derived from a fiducial
mass of 6.5~kg, defining a total `geometrical' exposure of
454~kg$\cdot$days, and a factor subsuming all energy-independent
hardware and software inefficiencies (and the restricted
acceptance). An energy-dependent detection efficiency, which reaches
unity near 14~keV recoil energy, is applied separately. Conversion
between visible and nuclear recoil energies utilises the varying
quenching factor (QF) discussed in Ref.~\cite{lebedenko08a}. The
exposure of the odd-neutron isotopes reflects their relative
abundance. Our xenon was depleted from high-mass isotopes, especially
$^{136}$Xe used in $0\nu\beta\beta$-decay experiments. This enhances
slightly the isotopic composition in $^{129}$Xe and $^{131}$Xe
relative to natural abundance (see Table~I), as confirmed by residual
gas analysis using mass-spectroscopy.

Seven events were observed in the acceptance region, all near the
upper boundary of the discrimination parameter. These were shown to be
statistically consistent with the tail of the electron-recoil
background population. To test this hypothesis in the presence of a
hypothetical WIMP signal, a maximum likelihood analysis was performed
on the acceptance region. This returned a most likely signal of zero
WIMP events, with 90\% upper limits increasing from 2.45 to 3.0 events
with increasing WIMP mass. The experimental limit on the WIMP-nucleus
cross-section, $\sigma_{A}$, was calculated as described
Ref.~\cite{lewin96}. We assumed an isothermal dark matter halo with
truncated Maxwellian velocity distribution with characteristic
velocity $v_0$=220~km/s, galaxy escape velocity $v_{esc}$=600~km/s,
Earth velocity $v_E$=230~km/s and WIMP mass density
$\rho_0$=0.3~GeV/cm$^3$. The limit on the differential recoil spectrum
was corrected by the normalised nuclear form factor and detector
parameters such as energy resolution and detection efficiency.

The SD cross-section limits, with no assumption on the coupling
strength to neutrons and protons, are shown in Figure~1 for the two
spin structures. Values at 55~GeV/$c^2$ WIMP mass, the minimum of the
curves, are given in Table~II. Our Bonn~A result surpasses that from
XENON10~\cite{angle08b} above 100~GeV (taking into account new QF
measurements for xenon~\cite{aprile08}). Together these experiments
place the world's most stringent limit on $\sigma_n$, when this $nn$
potential is assumed. However, Bonn~CD affects these limits
unfavourably: $\sigma_n$ is just under twice higher, but $\sigma_p$
increases by orders of magnitude, virtually eliminating the
sensitivity to WIMP-proton scattering (this curve is not shown in the
figure as it would fit poorly within the range plotted). Naturally,
these corrections apply to all xenon detectors. We note that
comparison with other experiments is not always straightforward: i)
different spin structures may be used for the same isotopes, as new
calculations are still emerging; ii) the model-independent approach of
Ref.~\cite{tovey00} is not used universally (e.g.~a combined SI and SD
limit is extracted in HDMS, where a novel background subtraction
technique is also employed~\cite{bednyakov08}) and iii) statistical
significance may vary, as with the XENON10 result, for which a
CL$<$90\% is ackowledged for the upper limit (c.f. note [26] in \cite{angle08a}).

\begin{figure}[t]
  \includegraphics[width=3.5in]{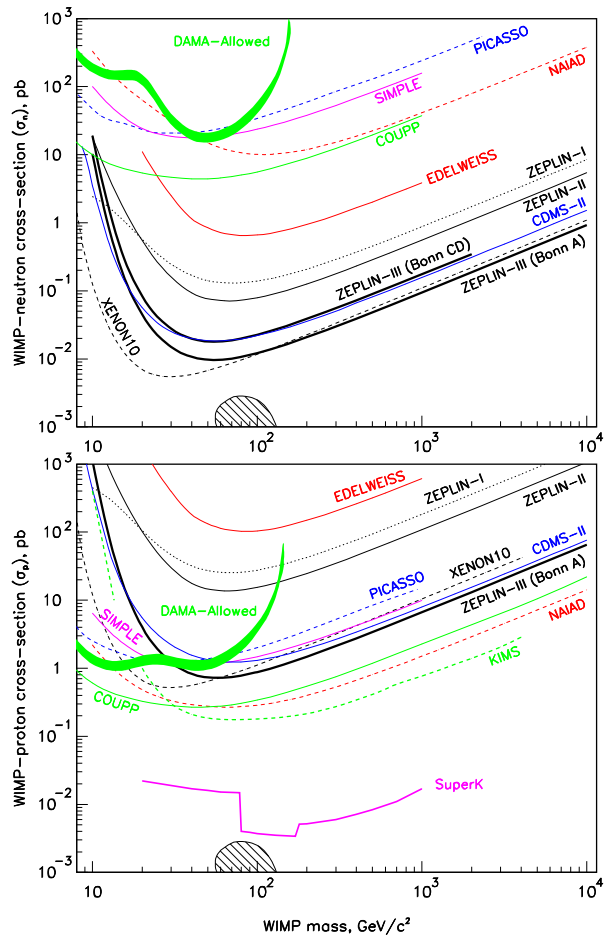}
  \caption{\label{fig1} Upper limits on pure WIMP-neutron and
    WIMP-proton SD cross-sections. In addition to ZEPLIN-III with both
    $nn$ potentials, we show other xenon experiments in black:
    ZEPLIN-I~\cite{alner04}, ZEPLIN-II~\cite{alner07} and
    XENON10~(Bonn A)~\cite{angle08b,aprile08}. Additional curves are
    CDMS-II~\cite{ahmed09}, COUPP~\cite{behnke08},
    EDELWEISS~\cite{benoit05}, KIMS~\cite{lee07},
    NAIAD~\cite{alner05}, PICASSO~\cite{barnabe05} and
    SIMPLE~\cite{giuliani04}. The pure-proton indirect limit from
    Super-Kamiokande is also shown~\cite{desai04}. The DAMA evidence
    region interpreted as a nuclear recoil signal in a standard halo
    \cite{savage04} is indicated in green. The hatched area is the tip
    of the 95\% probability region for neutralinos in the Constrained
    Minimal Supersymmetric Standard Model (CMSSM)
    \cite{roszkowski07}.}
\end{figure}

\begin{table*}
\caption{Spin-dependent cross-section limits (pb) at $M_W$=55~GeV for
$^{129}$Xe and $^{131}$Xe, and the combined ZEPLIN-III result.}
\begin{ruledtabular}
\begin{tabular}{lcccccc}
&$\sigma_n^{129}$ & $\sigma_p^{129}$ & $\sigma_n^{131}$ & $\sigma_p^{131}$ & $\sigma_n$ & $\sigma_p$ \\
\hline
Bonn A  & $1.2\!\times\!10^{-2}$ & $8.3\!\times\!10^{-1}$ & $5.7\!\times\!10^{-2}$ & $5.8\!\times\!10^{+0}$ & $9.7\!\times\!10^{-3}$ & $7.2\!\times\!10^{-1}$ \\
Bonn CD & $2.0\!\times\!10^{-2}$ & $4.8\!\times\!10^{+2}$ & $1.5\!\times\!10^{-1}$ & $2.0\!\times\!10^{+3}$ & $1.8\!\times\!10^{-2}$ & $3.9\!\times\!10^{+2}$
\end{tabular}
\end{ruledtabular}
\end{table*}

The allowed region of $a_n-a_p$ parameter space can be derived from
the experimental cross-section limits~\cite{tovey00}:
\begin{equation}
  \sum_i \left( \frac{a_p}{\sqrt{\smash{\sigma_p}^i}} \pm
  \frac{a_n}{\sqrt{\smash{\sigma_n}^i}} \right)^2
  \le \frac{\pi}{24G_F^2\mu_p^2} \,
  \label{ellipse}
\end{equation}
where $i$ labels the two Xe isotopes, and the sign in parenthesis is
that of $\langle S_p \rangle$/$\langle S_p \rangle$. For xenon and
other multi-isotope targets, (\ref{ellipse}) defines an ellipse --
albeit an elongated one -- which reduces to two parallel lines for
single-isotope experiments. The region allowed by each experiment lies
within the corresponding ellipse. Figure~2 shows our Bonn CD result
(nearly vertical lines, reflecting the poor constraint on $a_p$) for a
reference WIMP mass of 50~GeV/$c^2$ (the cross-sections coincide with
those in Table~II within the precision shown). The Bonn A result (not
shown) is very similar to that from XENON10.

\begin{figure}[t]
  \includegraphics[width=3.0in]{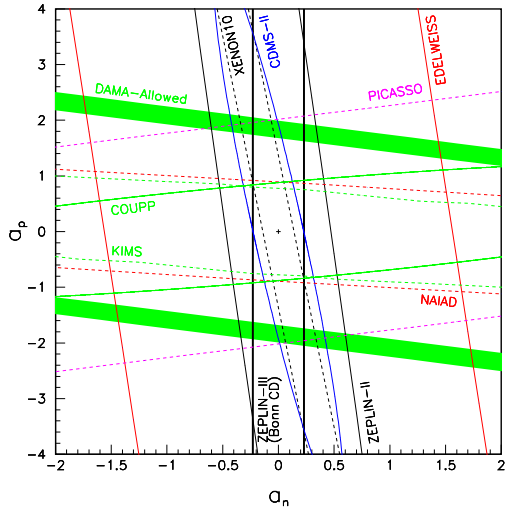}
  \caption{\label{fig:epsart} Allowed regions in the $a_n-a_p$
    parameter space for 50~GeV WIMP mass.}
\end{figure}


In conclusion, new experimental limits on the SD WIMP-nucleon
cross-section are placed by the first science run of ZEPLIN-III
operating at Boulby. A fiducial exposure of $\sim$450~kg$\cdot$days
revealed a most likely signal of zero events, which allow 90\% CL
exclusion of the pure WIMP-neutron cross-section above
$\sigma_n$=1.8$\times$10$^{-2}$~pb for 55~GeV/$c^2$ WIMP mass. New
spin structure calculations based on the Bonn CD $nn$ potential were
used for $^{129}$Xe and $^{131}$Xe. These increase the cross-section
limits relative to previous calculations -- quite dramatically in the
case of the proton. Xenon targets are still the very sensitive to the
WIMP-neutron interaction. Our result adds weight to the exclusion of
the DAMA evidence region for a 50~GeV/$c^2$ WIMP causing nuclear
recoils. Theoretical predictions in the CMSSM~\cite{roszkowski07}
point to $\tilde{\chi}$-nucleon cross-sections below
$\sim\!3\!\times\!  10^{-3}$~pb near 100~GeV mass at 95\%
CL. ZEPLIN-III, XENON10 and CDMS-II provide the leading constraint on
SD $\tilde{\chi}$-neutron scattering at that mass, nearly missing that
probability boundary. Experimental efforts are still far from probing
the favoured parameter space for the $\tilde{\chi}$-proton interaction
(the CMSSM best-fit regions suggest $\sim\!1\!\times\! 10^{-4}$~pb for
both nucleons).

This work was funded by the Science \& Technology Facilities Council
(UK), the Russian Foundation of Basic Research and the Portuguese {\it
Funda\c{c}\~{a}o para a Ci\^{e}ncia e Tecnologia}. The authors are
grateful to Cleveland Potash Ltd, who host the Boulby laboratory. We
also thank J.~Suhonen and V.~A~Kudryavtsev for useful discussions.

\bibliography{HAraujo}

\end{document}